\documentclass[a4paper]{article}
\usepackage{Odyssey2022}
\usepackage{epsfig,amssymb,amsmath,amsfonts}
\usepackage{algorithmic}
\ninept
\usepackage{array}
\usepackage[table]{xcolor}
\usepackage{multirow}
\title{Enhance Language Identification using Dual-mode Model with Knowledge Distillation}

\makeatletter
\def\name#1{\gdef\@name{#1\\}}
\makeatother
\name{{\em Hexin Liu$^1$, Leibny Paola Garcia Perera$^2$, Andy~W.~H.~Khong$^1$, Justin Dauwels$^3$}\\
     {\em  Suzy~J.~Styles$^4$, Sanjeev Khudanpur$^2$}}

\address{$^1$School of Electrical and Electronic Engineering, Nanyang Technological University, Singapore\\
  $^2$CLSP and HLT-COE, Johns Hopkins University, USA\\
  $^3$Department of Microelectronics, Delft University of Technology, Netherlands\\
  $^4$Psychology, School of Social Sciences, Nanyang Technological University, Singapore\\
{\small \tt HEXIN002@e.ntu.edu.sg, lgarci27@jhu.edu} }
\begin{document}
\maketitle

\begin{abstract}
In this paper, we propose to employ a dual-mode framework on the x-vector self-attention~(XSA-LID) model with knowledge distillation~(KD) to enhance its language identification~(LID) performance for both long and short utterances. The dual-mode XSA-LID model is trained by jointly optimizing both the full and short modes with their respective inputs being the full-length speech and its short clip extracted by a specific Boolean mask, and KD is applied to further boost the performance on short utterances. In addition, we investigate the impact of clip-wise linguistic variability and lexical integrity for LID by analyzing the variation of LID performance in terms of the lengths and positions of the mimicked speech clips. We evaluated our approach on the MLS14 data from the NIST 2017 LRE. With the 3~s random-location Boolean mask, our proposed method achieved 19.23\%, 21.52\% and 8.37\% relative improvement in average cost compared with the XSA-LID model on 3~s, 10~s, and 30~s speech, respectively. 
\end{abstract}

\section{Introduction}
\label{sec:intro}

Spoken language identification~(LID) refers to the automatic process through which we determine the identity of the language spoken in a speech sample \cite{li2013spoken}. Conventional LID methods such as the i-vector and time-delay neural network-based~(TDNN) x-vector methods are based on a two-stage process \cite{dehak2011language, snyder2018spoken}\textemdash language embeddings are first extracted from the speech using a front-end encoder, and an independently trained classifier is then employed back-end to identify the language using the embeddings. Recently, the widely investigated end-to-end methods integrate this two-stage process into a single neural module \cite{cai2019utterance, Miao2019, wan2019tuplemax,ling20_odyssey}.

Although existing methods have shown promising results on general LID tasks, achieving LID on short utterances is still challenging. Recent deep neural network-based approaches have shown to be more effective than conventional i-vector approaches for short utterances \cite{Miao2019, framebyframe14}. For instance, Lozano-Diez et al.~proposed to train a convolution deep neural network~(CDNN)-based mechanism on short utterance, i.e.,~3~s speech \cite{LozanoDiez2015AnEA}. A bidirectional long short-term memory~(BLSTM) network has also been proposed to model the temporal dependencies between the past and future frames in short utterances \cite{fernando2017bidirectional}. In addition, Peng et al. proposed several methods based on knowledge distillation~(KD) or compensation on x-vector to transfer the knowledge of a pre-trained long-utterance LID model to a short-utterance LID model \cite{shen2018feature, interactiveL, compensation2020}. 

These approaches, however, focus primarily on short utterances and are not developed to achieve good performance on long utterances. Practical applications of the LID system, however, should cater to speech of varying duration. Moreover, some existing methods require an additional data preprocessing step such as speech segmentation and a pre-trained model or additional parameters. These components are necessary to facilitate the distillation of the knowledge from the model trained on long utterances to the target short-utterance LID model \cite{shen2018feature, interactiveL, compensation2020}.

Aside from LID, streaming automatic speech recognition~(ASR) aims to generate each hypothesized word as quickly and accurately as possible without the availability of future frames \cite{streamMobile, dualasr}. In \cite{dualasr}, a unified dual-mode ASR model has been proposed to improve the streaming ASR with full-context modelling, in which the streaming ASR model shares the same parameters with the full-context ASR model and the future information is masked in the streaming mode.

Challenges posed by the availability of only a portion of speech in streaming ASR is similar to that of short-utterance LID. Inspired by the success of the dual-mode ASR system in streaming ASR, we propose a dual-mode LID model based on KD and the x-vector self-attention end-to-end~(XSA-E2E) model to enhance the LID performance on speech with various duration \cite{hinton2015distilling, liuxsa}. Apart from the performance enhancement in LID, our proposed dual-mode LID approach possesses several desirable properties. Firstly, the proposed method employs the KD method without introducing an additional model or parameters. Secondly, instead of preprocessing using speech segmentation, a Boolean mask is applied to acquire features more flexibly with lower computational complexity. In addition, different from the use of mask in dual-mode ASR which aims to remove the future frames from the streaming mode, the masking operation in the proposed model mimics short speech clips. The various types of linguistic information introduced by the mask serves as data augmentation and leads to higher LID performance. Lastly, since the Boolean mask is applied to the transformer encoder layers in the proposed model, our approach can be easily transferred.

The remainder of this paper is organized as follows: Section~\ref{sec:relate} introduces the work related to this paper. The XSA-LID model and our proposed dual-mode LID model are illustrated Section~\ref{sec:method}. The experiment setup are then given in Section~\ref{sec:experiment}. We present the results and analysis including the evaluation of different systems and the comparison between different Boolean masks in Section~\ref{sec:results}. We propose the strategies for our proposed system to achieve high performance on the target data in terms of duration in Section~\ref{sec:discussion}, and conclude our work in Section~\ref{sec:conclusion}.

\section{Related work}
\label{sec:relate}

In the dual-mode ASR system \cite{dualasr}, the full-context ASR model applies a Boolean mask to mimic the streaming ASR model. These models share the same weights and are jointly trained with connectionist temporal classification~(CTC) and KD losses. In contrast to the dual-mode ASR model which focuses more on the streaming ASR task, our work aims to improve the LID performance on both long and short utterances. Therefore, we adapt the dual-mode framework to the LID task by modifying the mask to flexibly acquire linguistic information. In addition, considering the success of KD in both short-utterance LID and streaming ASR \cite{interactiveL,dualasr}, we adopt the KD method to improve the LID performance on short utterances.

The XSA-E2E model was recently proposed for language diarization \cite{liuxsa}. The model employs an x-vector embedding module for 200~ms segment-level encoding, and the transformer encoder module is then applied to learn the global dependencies of these segment-level embeddings. Since the diarization process comprises segmentation and identification, segment-level LID naturally exists in language diarization. Therefore, we propose to modify the architecture to adapt the XSA-E2E model to the LID task. This adapted LID model is denoted as XSA-LID. 

\section{Methodology}
\label{sec:method}
\subsection{Adaptation of the XSA-E2E model for LID}
\label{sec:xsa}
\begin{figure}[t]
  \centering
  \includegraphics[width=\linewidth]{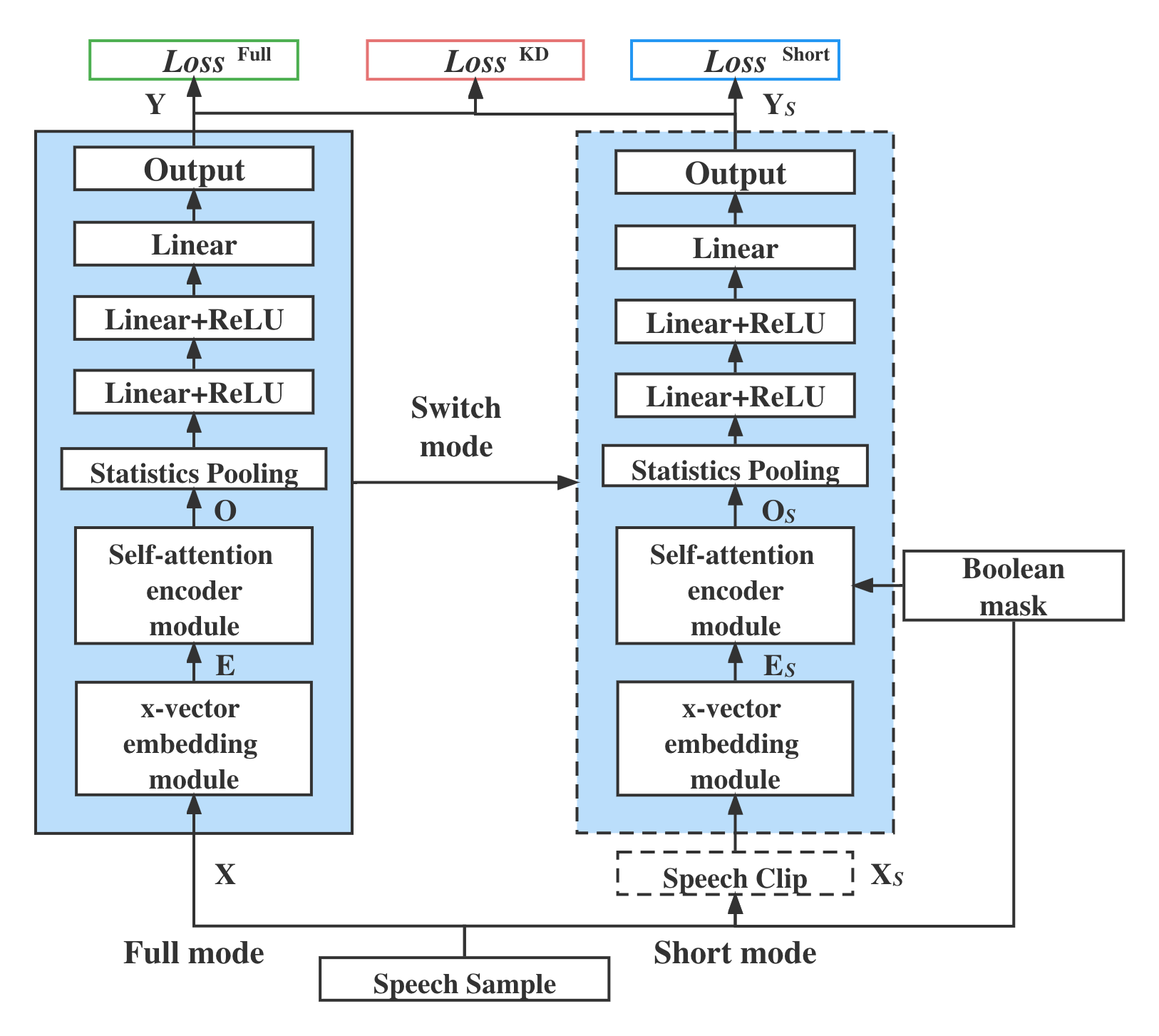}
  \caption{Dual-mode LID system with X-vector-Self-Attention LID~(XSA-LID) model and knowledge distillation.}
  \label{fig:xsae2e}
\end{figure}

With reference to the full mode shown in Fig.~\ref{fig:xsae2e}, we adapt the XSA-E2E model for LID and denote it as XSA-LID. We define \(\mathbf{X} = (\mathbf{x}_{t}\in\mathbb{R}^{K\times F}| t=1, ... , T)\) as input of the XSA-LID model, where \(\mathbf{X}\) comprises features extracted from the input speech signal partitioned in segments, and \(T\) is the number of segments. Each segment \(\mathbf{x}_{t}\) comprises \(K\) frames \(\left [ \mathbf{f}_{1},...,\mathbf{f}_{K} \right ]^{\intercal}\), where \(\mathbf{f}_{k}\) is an \(F\)-dimensional frame-level feature vector of the \(k\)-th frame in segment \(t\), and the original speech signal consists of \(TK\) frames.

Similar to the XSA-E2E model, the front-end x-vector embedding module in XSA-LID model comprises TDNN layers and a statistic pooling layer followed by a linear layer. This module captures the local language information in every segment \(\mathbf{x}_{t}\) and generates the corresponding segment-level embeddings \(\mathbf{E} = (\mathbf{e}_{n} \in \mathbb{R}^{D}| n=1, ... , T)\) which are subsequently fed into the self-attention encoder module. 
The self-attention encoder module consists of the transformer encoder layers and learns the global dependencies of the segment-level embeddings. Defining $D$ as the output dimension of the transformer encoder layers, the output \(\mathbf{O} = (\mathbf{o}_{n} \in \mathbb{R}^{D}| n=1, ... , T)\) is computed as
\begin{equation}
  \mathbf{O}= \mathrm{SelfAtten}\left ( \mathbf{E}, \mathbf{Mask}\right ),
  \label{eq:self_atten_module}
\end{equation}
where \(\mathrm{SelfAtten}\left ( \cdot,\cdot  \right )\) denotes computations performed within the self-attention encoder module, and \(\mathbf{Mask}\) is defined as the attention mask applied to the attention weight matrix, i.e., the dot product between the query and key matrices \cite{vaswani2017attention}. In our implementation, given a batch of training samples, the utterances (which are shorter than the longest one) are padded to be of the equal length so that they can be fed into the model. These padded components in the matrix corresponding to the ``False" in \(\mathbf{Mask}\) are set to a significantly low value such as \(-10^{9}\). Therefore, these components will not contribute to the subsequent computations after the softmax function.

A statistics pooling layer is employed to generate an utterance-level representation for the input speech signal. This is achieved by aggregating the segment-level output \(\mathbf{O}\) of the encoder module. The following fully connected layers generate class scores \(\mathbf{Y}\) of the target languages. Finally, during the training phase, the XSA-LID model is optimized by minimizing the CE loss instead of the joint loss proposed in \cite{liuxsa}. Details of the configuration will be described in Section~\ref{sec:experiment}. 

\subsection{Dual-mode language identification}
\label{sec:dual}
\subsubsection{Dual-mode framework}
To enhance the performance of the XSA-LID model, we next adopt the dual-mode architecture, which was originally proposed for streaming ASR. With reference to Fig.~\ref{fig:xsae2e}, during training, the full mode corresponds to the XSA-LID model with the input full-length speech being \(\mathbf{X}\). On the other hand, since features corresponding to the mimicked short speech clip of the full-length speech are extracted from $\mathbf{X}$ via a Boolean mask applied in Equation~\ref{eq:self_atten_module}, the input of the short mode is also \(\mathbf{X}\). Here, it is equivalently given by \(\mathbf{X}_{S} = (\mathbf{x}_{t}\in \mathbb{R}^{K\times F}| t=s, ... , s+{T}_{S}-1)\), where $T_S<T$ is the number of segments in the speech clip. We jointly optimize these two modes in our proposed dual-mode LID model. The full mode, therefore, achieves general LID, while the short mode focuses on the local information in the speech clips. Hence, the dual-mode method can achieve good LID performance on both full-length speech and short speech. 

To achieve good system performance for the LID task without the use of additional dataset, one possible choice is to train the system with the data augmentation using speech clips segmented from the full-length speech. While the model trained with such augmentation can achieve good performance on short utterances, having larger number of short clips in the updates generally results in performance degradation on long utterances.

As opposed to data augmentation using short speech clips, features of the mimicked short speech clips in the short mode of our proposed model is being represented by the full-length speech in the full mode in the current update. Since the short mode better grasps a portion of local information of the full-length speech and the utterance-level embedding is generated by aggregating the segment-level statistics in the statistics pooling layer, the enhanced ability to capture local language information results in more improvement in the performance on long utterance compared to short utterances. Consequently, by minimizing the sum of the losses of these two modes, the proposed model is able to achieve significant performance improvement on long utterances with a modest improvement on short utterances. 
\begin{figure}[t]
  \centering
  \includegraphics[width=60mm]{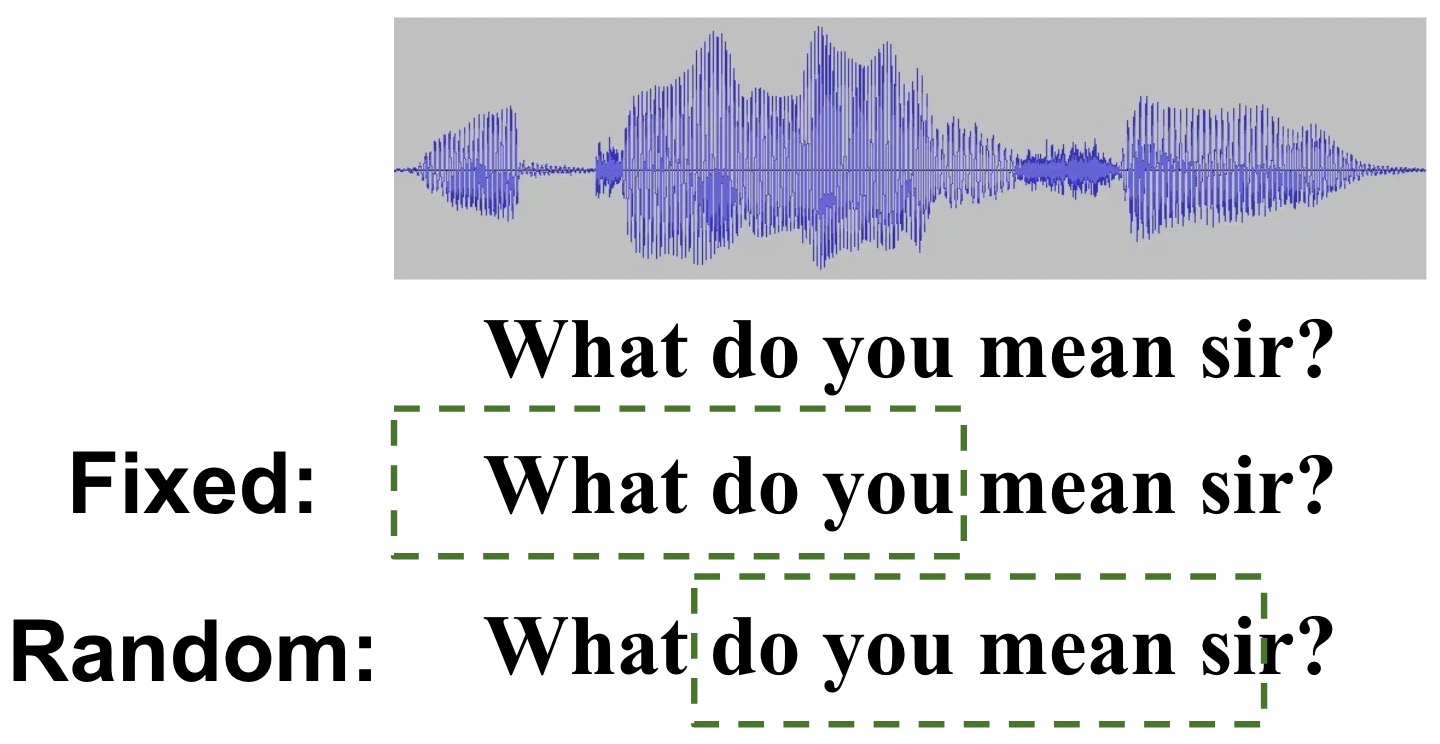}
  \caption{Two types of Boolean mask for acquiring linguistic information. The text bounded by the green dotted lines corresponds to the mimicked speech clip}
  \label{fig:bool_mask_linguistic}
\end{figure}
\subsubsection{Knowledge distillation loss}
\label{sec:kd}
To improve the short-utterance LID performance on the dual-mode framework, we employ the KD loss that minimizes the difference between prediction probability distributions of the full-length speech and the corresponding short clip. The KD loss possesses two important characteristics for LID. Firstly, compared to the CE loss in the dual-mode framework, the KD loss utilizes the output of the full mode which incorporates valuable information pertaining to non-target languages, e.g., dialects. This allows the short mode to learn rich discriminative information from the full mode. Secondly, due to the minimization of the difference between outputs of the two modes, the introduction of KD loss to the optimization may reduce the LID performance for long utterances. However, this performance degradation for long utterances can be offset by the improvement resulted by the dual-mode framework which is optimized via CE losses. Therefore, the proposed model can achieve higher performance on speech of varying duration due to the trade-off between KD loss and CE losses.

It is useful to note that two independent utterances can differ from each other in terms of language-unrelated identities such as speaker and lexical information. These identities, however, affect the LID performance. Our approach applies the KD loss to the full-length speech and its short clip, which can reduce the effect of duration-unrelated information. Thus, the KD loss which focuses on the difference of time duration is computed via~\cite{hinton2015distilling}
\begin{eqnarray}
  P\left(\mathbf{X}\right) &=& \mathrm{Softmax}\left( \frac{\mathbf{Y}}{\mathbf{Temp}}\right),
  \label{eq:loss_soft}
  \\
  L^{\mathrm{KD}} &=& \mathrm{KL}\left ( \mathrm{log}\left( P\left(\mathbf{X}\right)\right), P\left( \mathbf{X}_{S}\right) \right ),
  \label{eq:loss_kd}
\end{eqnarray}
where \(\mathbf{Y}\) denotes the system outputs, \(\mathbf{Temp}\) is the distillation temperature, \(\mathrm{Softmax}\left ( \cdot  \right )\) is the softmax operation, \(\mathrm{KL}\left ( \cdot,\cdot  \right )\) denotes the KL divergence, and \(P\left ( \cdot  \right )\) denotes the output probability for language classes given input speech features. During training, our proposed model is optimized by minimizing the weighted sum of KD loss and the CE losses of the full-mode and short-mode. This loss is given by
\begin{equation}
  L^{\mathrm{Dual}} = \alpha L^{\mathrm{Full}} + \beta L^{\mathrm{Short}} + \left ( 1-\alpha-\beta  \right )L^{\mathrm{KD}},
  \label{eq:loss_all}
\end{equation}
where \(\alpha\) and \(\beta\) are parameters to compensate for any trade-off in performance between short and long utterances. In particular, to achieve a balanced performance on long and short utterances, \(\alpha\), \(\beta\) and \(\left ( 1 -\alpha -\beta \right )\) are set to be equal. Since the performance improvement on short utterances is mainly attributed to the KD loss, higher \(\beta\) and \(\left ( 1 -\alpha -\beta \right )\) are adopted in the presence of frequently occurring short utterances in the target data. Conversely, higher \(\alpha\) and \(\beta\) will be required to achieve high LID performance for long utterances.
\begin{figure}[t]
  \centering
  \includegraphics[width=55mm]{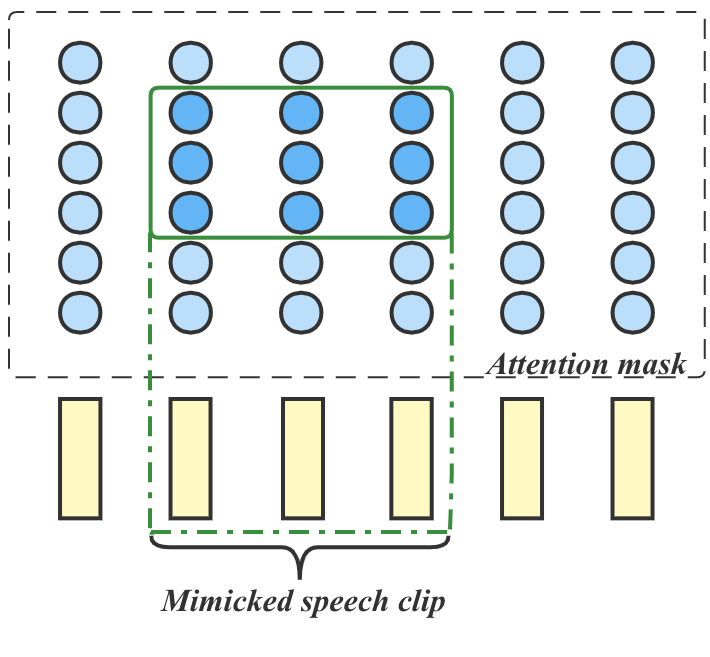}
  \caption{The random-location Boolean mask for the self-attention encoder module. In this figure, the mimicked short speech clip starts from the second segment to the fourth segment. The shallow blue circles are filled with a significantly small value before the softmax operation when computing the attention weight matrix}
  \label{fig:boolean_sa_enc}
\end{figure}
\subsubsection{The Boolean mask for mimicking short speech clips}
\label{sec:booleanmask}
To mimic the short speech clip and acquire various clip-wise linguistic information, with reference to Equation~\ref{eq:self_atten_module}, we utilize the Boolean mask, which was originally applied to the transformer encoder layers of the XSA-LID model, to handle input utterances of varying duration. In contrast to the Boolean mask in the dual-mode ASR which masks future frames, we propose two types of masking strategies for short-utterance LID: 1) the fixed-location mask through features of the first several seconds speech of the full-length speech are extracted and 2) the random-location mask through which features of a specified window duration from a random time step of the full-length speech are extracted.

To illustrate the above, we consider the utterance ``What do you mean sir" shown in Fig.~\ref{fig:bool_mask_linguistic}. Since it is higher likely that the random-position masking cuts off the voiced speech compared to the fixed-position masking, the speech clip ``What do you" mimicked by the fixed-location mask contains more clip-wise lexical integrity than the clip ``do you mean si" mimicked by the random-location mask. Nonetheless, the random mask can derive different short clips for each utterance over the training epochs, while the fixed mask always locates in the first several seconds. Therefore, the random mask is expected to introduce richer clip-wise linguistic variability for training, while the fixed mask provides better clip-wise lexical integrity to the mimicked short clips. In addition, the model with a random-location mask is expected to achieve higher performance on short utterances since richer short-duration information is exploited for the training, while the model trained with better clip-wise lexical integrity in the fixed-location mask may exhibit higher improvement for long utterances.

We illustrate the random-location Boolean mask in Fig.~\ref{fig:boolean_sa_enc}, where the components circled by the green bounding box correspond to the mimicked short clip. Considering that the x-vector module in the XSA-LID model represents segments as vectors, no modification was made to the TDNN layers. The Boolean mask is then employed in the self-attention computation. By replacing the \(\mathbf{Mask}\) in Equation~\ref{eq:self_atten_module} in the full mode with this Boolean mask, segment-level embeddings which are unrelated to the speech clip are filtered out\textemdash the remaining segment-level embeddings therefore correspond to the mimicked speech clip. Optimization with the Boolean mask is shown as the pseudocode of the dual-mode LID in Table~\ref{tab:algorithm}.

\section{Dataset, Experiment, and Model Configuration}
\label{sec:experiment}
\subsection{Dataset and feature extraction}
In terms of datasets used for performance evaluation, the NIST LRE 2017 dataset consists of Fisher corpus, Switchboard corpora, a narrow-band telephony training dataset (TRN17) built from previous LRE data with over 2000 hours of audio data, a development dataset (DEV17), and an evaluation dataset (EVAL17) \cite{lre17_odyssey,sadjadi18_interspeech}. The DEV17 and EVAL17 comprise narrow-band MLS14 data and wide-band VAST data, while the MLS14 data consist of 3~s, 10~s and 30~s three different duration levels and the VAST data consist of segments with speech duration ranging from 10~s to 600~s. 

\begin{table}[t]
\centering
\resizebox{.47\textwidth}{!}{
\begin{tabular}{lll}
\hline
 \multicolumn{3}{l}{\textbf{Algorithm:} Pseudocode of the dual-mode language identification}     \\ \hline
 \multicolumn{3}{l}{\textcolor{blue!50!}{\# Load a batch of inputs speech features x and language labels y}}    \\
 \multicolumn{3}{l}{for x, y in data\_loader: }                                          \\
 & \multicolumn{2}{l}{\textcolor{blue!50!}{\# full-mode: inputs are full-length x}} \\
 & \multicolumn{2}{l}{with dual\_mode\_model.mode("full"):}                  \\
 &            & \textcolor{blue!50!}{\# compute full-length prediction given x and y }\\
 &            & full\_pred = dual\_mode\_model.forward(x, y, atten\_mask) \\
 &            & \textcolor{blue!50!}{\# compute cross-entropy loss for full-mode }\\
 &            & loss\_full = CE(full\_pred, y)               \\
 & \multicolumn{2}{l}{\textcolor{blue!50!}{\# short-mode: inputs are clips of x, one clip for each sample in x }} \\
 & \multicolumn{2}{l}{with dual\_mode\_model.mode("short"):} \\
 &            & \textcolor{blue!50!}{\# built a Boolean mask give the length of the speech clip }\\
 &            & short\_mode\_mask = attention\_mask(speech\_clip\_length)\\
 &            & \textcolor{blue!50!}{\# compute short-utterance prediction given x, y and the mask }\\
 &            & short\_pred = dual\_mode\_model.forward(x, y, short\_mode\_mask)\\
 &            & \textcolor{blue!50!}{\# compute cross-entropy loss for short-mode }\\
 &            & loss\_short = CE(short\_pred, y)\\
 & \multicolumn{2}{l}{\textcolor{blue!50!}{\# compute the KD loss, teacher is the full-mode }} \\ 
 & \multicolumn{2}{l}{\begin{tabular}[c]{@{}l@{}}loss\_KD = knowledge-distillation-loss(full\_pred, short\_pred) \\
 \textcolor{blue!50!}{\# compute the dual-mode loss and optimize the model }\\
 loss\_dual\_mode = \(\alpha\) loss\_full + \(\beta\) loss\_short + \((1-\alpha-\beta)\) loss\_KD\\     
 loss.backward() \end{tabular}} \\ \hline
\end{tabular}}
\label{tab:algorithm}
\end{table}

To ensure fair evaluation, we have used the same features for all systems. We extracted 80-dimensional bottleneck features~(BNFs) from an ASR-DNN model that is trained on the Fisher corpus and Switchboard corpora \cite{cieri2004fisher,godfrey1992switchboard}. The input features of the ASR-DNN model are 13-dimensional Mel frequency cepstral coefficients (MFCCs) extracted from a 25~ms window with a 10~ms shift. Silent frames are removed using an energy-based voice activity detection. 
We trained all systems on TRN17 and DEV17 and tested them on the MLS14 data in EVAL17 to compare their performance on the speech of different duration levels. To reduce the length of frame sequences for the transformer and conformer encoders, every 20-frame BNFs are concatenated into a new unit and the 1600-dimensional BNFs are subsequently fed into these two models during both training and inference. Feature extraction is performed using the Kaldi toolkit \cite{Povey_ASRU2011}.
\subsection{Model configuration}
In terms of the configuration of systems in Table~\ref{tab:lre17}, the x-vector approach follows that presented in \cite{snyder2018spoken} and is trained by modifying the SRE16 recipe in the Kaldi toolkit with a back-end logistic regression classifier. The transformer and conformer models refer to the encoder blocks presented in \cite{vaswani2017attention, gulati2020conformer}. These models consist of four self-attention encoder layers followed by a statistics pooling layer and three linear layers with ReLU activation in the first two linear layers.

As shown in Fig~\ref{fig:xsae2e}, the XSA-LID model comprises an x-vector embedding module followed by the self-attention encoder module, an utterance-level statistic pooling layer and three linear layers with ReLU activation in the first two linear layers. The x-vector embedding module consists of three TDNN layers followed by a statistics pooling layer and a linear layer. The input dimension \(F\)=80 corresponds to the dimension of the BNFs. The TDNN layers are conv1d layers with kernel size \(\left (5,5,1 \right )\), dilation \(\left (1,2,1 \right )\) and output dimensions \(\left (512,512,512 \right )\). These layers generate an embedding for each segment of duration twenty frames with each segment being approximately 200~ms. The linear layer projects the 1024-dimensional output of the segment-level statistics pooling layer to 64-dimensional. Two transformer encoder layers in the self-attention encoder module follow \cite{vaswani2017attention}, each encoder layer has eight attention heads with \(d_{model}\)=512 and \(d_{ff}\)=2048. The utterance-level statistics pooling layer then generates the 1024-dimensional output which is finally projected by three linear layers to the number of target languages. The output dimensions of these three linear layers are \(\left (512,512,14 \right )\).

Our proposed dual-mode LID model is based on the XSA-LID model. Variables $\alpha=\beta=0.33$ in Equation~\ref{eq:loss_all} were chosen to assign nearly equal importance to the two modes and KD loss. A 3~s Boolean mask is applied in our work, which covers 15 segments. The distillation temperature is set to 2. These NN-based models are all trained for 20 epochs using the Adam optimizer with an initial learning rate of \(10^{-4}\) and cosine annealing learning rate decay after 24000 warm-up steps. A batch size of 32 is use for the transformer, conformer and the XSA-LID models, and 16 for our proposed dual-mode LID model.
\subsection{Performance measurement}
We evaluated our systems by employing accuracy (Acc.), equal error rate (EER) and the average cost (\({C}_{avg}\)) \cite{lre07}. Since the EER and accuracy are widely applied in other areas such as speaker recognition, we explain how \({C}_{avg}\) is computed.
\begin{table*}[t]
\centering
\caption{Results of different models on the MLS14 data in NIST LRE 2017 by employing Accuracy~(\%), EER~(\%) and \({C}_{avg}\) and ablation study}
\label{tab:lre17}
\setlength{\tabcolsep}{1.6mm}{
\begin{tabular}{|c|c|c|c|c|c|c|c|c|c|c|c|c|}
\hline
\multirow{2}{*}{\textbf{Method}}      & \multicolumn{3}{c|}{3~s}      & \multicolumn{3}{c|}{10~s}              & \multicolumn{3}{c|}{30~s}   & \multicolumn{3}{c|}{Overall Avg.}  \\ \cline{2-13}
& Acc. & EER  & \({C}_{avg}\) & Acc. & EER  & \({C}_{avg}\) & Acc. & EER & \({C}_{avg}\)  & Acc. & EER & \({C}_{avg}\) \\ \hline
X-vector \cite{snyder2018spoken}    
& -              & \textbf{11.79}          & \textbf{0.1159}          & -             & 7.81          & 0.0748          & -              & 6.84        & 0.0654  & -     & 8.81  & 0.0854\\ \hline
Transformer encoder \cite{vaswani2017attention} 
& 44.47          & 19.66          & 0.1998          & 69.60         & 9.00          & 0.0792          & 79.58          & 5.81        & 0.0487  & 64.55 & 11.49 & 0.1123 \\
Conformer encoder \cite{gulati2020conformer}  
& 49.29          & 16.84          & 0.1627          & 73.30         & 7.76          & 0.0684          & 82.57          & 4.83        & 0.0393  & 68.39 & 9.81  & 0.0901\\ \hline
XSA-LID \cite{liuxsa}     
& 54.18          & 15.47          & 0.1685          & 74.90         & 7.39          & 0.0739          & 84.09          & 4.46        & 0.0406  & 71.06 & 9.11  & 0.0943    \\
XSA-Aug     
& \textbf{58.28} & 13.23          & 0.1308          & 75.28         & 7.32          & 0.0674          & 81.05          & 5.29         & 0.0470 & 71.53 & 8.61  & 0.0817         \\
DualNoKD    
& 55.35          & 15.94          & 0.1648          & 77.17         & 6.51          & 0.0641          & 86.13          & \textbf{3.88} & \textbf{0.0370} & 72.88 & 8.78 & 0.0886\\
Dual-LID-Fixed    
& 57.00          & 14.30          & 0.1420.         & 78.97         & 6.46          & 0.0595          & \textbf{87.02} & 4.00        & 0.0372  & 74.33 & 8.25  & 0.0796 \\ 
Dual-LID-Random   
& 58.25  & 13.82  & 0.1361    & \textbf{80.08} & \textbf{6.25} & \textbf{0.0580} & 86.65 & 4.09  & 0.0372 & \textbf{74.99} &\textbf{8.05} & \textbf{0.0771}\\ 
\hline
\end{tabular}}
\end{table*}

There are two types of errors in the LID task. For instance, for the LID task of \(Q\) languages, each language can be seen as a target language \(L_{T}\) and the other \(Q-1\) languages are non-target languages denoted as \(L_{N}\). The first type of error is defined as the miss error and occurs when the model misclassifies the target language as non-target. The second type of error is known as false alarm when the model misclassifies an impostor (non-target language) as the target. Pair-wise recognition performance is computed for all target/non-target language pairs by employing the miss and false alarm~(FA) rates such that a single cost performance is defined as
\begin{equation}
\begin{split}
  C\left ( L_{T},L_{N} \right )=&\mathrm{C_{miss}}\cdot \mathrm{P_{Target}}\cdot P_{miss}\left ( L_{T} \right )+\\
  &\mathrm{C_{FA}}\cdot \left ( 1-\mathrm{P_{Target}} \right )\cdot P_{FA}\left(L_{T},L_{N}\right),
\label{cost}
\end{split}
\end{equation}
where \(\mathrm{C_{miss}}=\mathrm{C_{FA}}=1\) and \(P_{Target}=0.5\) are predefined parameters, \(P_{miss}\left ( L_{T} \right )\) denotes the miss rate of \(L_{T}\), and \(P_{FA}\left(L_{T},L_{N}\right)\) denotes the false alarm rate given \(L_{T}\) and \(L_{N}\). The average cost is finally computed via
\begin{equation}
\begin{split}
  C_{avg}=&\frac{1}{\mathrm{Q}}\sum_{L_{T}}\left.\{ \mathrm{C_{miss}}\cdot \mathrm{P_{Target}}\cdot P_{miss}\left ( L_{T} \right )+\right.\\
  &\left.\sum_{L_{N}}\mathrm{C_{FA}}\cdot \left ( 1-\mathrm{P_{Target}} \right )\cdot P_{FA}\left(L_{T},L_{N}\right) \right.\}.
\label{cavg}
\end{split}
\end{equation}

\section{Results}
\label{sec:results}

\subsection{Results of different LID systems and ablation study}
\subsubsection{Results of different LID systems}
Results of different LID systems evaluated on the MLS14 data in EVAL17 are shown in Table~\ref{tab:lre17} across three different test speech durations. We have also included an ablation study pertaining to our proposed dual-mode LID system. Since the number of test speech across three durations are equal, we average their results to provide an indication of the overall performance. 

Compared to other approaches, the x-vector model exhibits the highest performance on 3~s speech but the lowest performance on 30~s utterances. It is not surprising that the transformer encoder, conformer encoder, and XSA-LID models show higher performance on 30~s test speech against the TDNN-based x-vector model since the transformer-based models have been proven to be successful in modeling long-term dependency. Our proposed XSA-LID model achieves the highest performance among these three models. This suggests that the ability to capture local information provided by the x-vector embedding module helps improve the LID performance. In addition, due to the superior performance of the x-vector model on 3~s speech, although the XSA-LID model achieves higher performance on 10~s and 30~s test speech, the x-vector model shows higher overall performance.

With regards to our proposed Dual-mode LID method, we denote the model with 3~s speech clips mimicked by random-location and fixed-location masks as Dual-LID-Random and Dual-LID-Fixed, respectively. The Dual-LID-Random model achieves 19.23\%, 21.52\%, and 8.37\% relative improvement in \(C_{avg}\) on 3 s, 10 s, and 30 s speech, respectively, compared with the XSA-LID model, and the best overall performance among all systems in terms of metrics under consideration.
\subsubsection{Ablation study}
To analyze the effects of the short clips, dual-mode structure and KD loss, we performed ablation studies using the results of XSA-Aug and DualNoKD systems. The XSA-Aug is the XSA-LID model which is trained with data augmentation by the first 3~s speech clips of utterances in TRN17 and DEV17. Therefore, XSA-Aug utilizes the same data as the Dual-LID-Fixed without the dual-mode framework nor the KD loss. We note from Table~\ref{tab:lre17} that XSA-Aug exhibits higher performance on 3~s and 10~s speech than the XSA-LID model. However, although the XSA-Aug system achieves higher performance on 3~s speech performance than the Dual-LID-Fixed, its performance on longer speech, especially for 30~s speech, is lower. This indicates that the augmentation by short speech clips, when enhancing the performance on 3~s and 10~s utterances, degrades the performance on 30~s speech;. In addition, secondly, these results show that the improvement of LID performance is not attributed only to the data but also by our proposed dual-mode LID method.

We next verify the importance of the dual-mode framework and KD loss using the DualNoKD model\textemdash the same model as the Dual-LID-Random model without KD loss. Compared to the XSA-LID model, the DualNoKD system achieves 2.2\%, 13.26\% and 8.87\% relative improvement in \({C}_{avg}\) on 3~s, 10~s and 30~s test data, respectively. This is consistent with our assumption described in Section~\ref{sec:dual} that the dual-mode framework can achieve more performance improvement on long utterances than short utterances. 

We also note from the above results that, after applying the KD loss to the DualNoKD system\textemdash the Dual-LID-Random model achieves 17.42\%, 9.52\% relative improvement in \({C}_{avg}\) on the 3~s and 10~s speech utterances, respectively, albeit suffering from modest performance degradation on 30~s test speech compared to the DualNoKD model. This accords with our supposed characteristics of the KD loss being applied to the dual-mode framework.

\begin{table*}[t]
\centering
\caption{Comparison of the performance of dual-mode LID systems with different Boolean masks by employing Accuracy~(\%), EER~(\%) and \({C}_{avg}\)}
\label{tab:mask_len_type}
\setlength{\tabcolsep}{3.5mm}{
\begin{tabular}{|c|c|c|c|c|c|c|c|c|c|c|}
\hline
\multicolumn{2}{|c|}{\textbf{Mask}} & \multicolumn{3}{c|}{3s} & \multicolumn{3}{c|}{10s} & \multicolumn{3}{c|}{30s} \\ \hline
\textbf{Location} & \textbf{Length} & Acc. & EER & \(C_{avg}\) & Acc. & EER & \(C_{avg}\) & Acc. & EER & \(C_{avg}\) \\ \hline
\multirow{6}{*}{Random}
& 1s & 56.89 & 14.36 & 0.1399 & 79.17 & 6.51 & 0.0601 & 86.77 & 4.24 & 0.0394 \\ \cline{2-11}
& 2s & 57.10 & 14.13 & 0.1409 & 78.63 & 6.43 & 0.0596 & 86.14 & 4.34 & 0.0402 \\ \cline{2-11}
& 3s & \textbf{58.25} & \textbf{13.82} & \textbf{0.1361} & 80.08 & \textbf{6.25} & 0.0580 & 86.65 & 4.09 & 0.0372 \\ \cline{2-11}
& 4s & 57.46 & 14.17 & 0.1376 & 79.25 & 6.38 & 0.0573 & 87.22 & 4.34 & 0.0369 \\ \cline{2-11}
& 5s & 57.95 & 14.20 & 0.1422 & \textbf{80.11} & 6.28 & 0.0575 & \textbf{87.24} & 4.24 & \textbf{0.0362} \\ \cline{2-11}
& 6s & 57.84 & 14.15 & 0.1391 & 79.51 & \textbf{6.25} & \textbf{0.0557} & 87.03 & 4.19 & 0.0376 \\ \hline
Fixed & 3s & 57.00 & 14.30 & 0.1420 & 78.97 & 6.46 & 0.0595 & 87.02 & \textbf{4.00} & 0.0372 \\ \hline
\end{tabular}}
\end{table*}

\subsection{Results of the LID models with different Boolean masks}
\subsubsection{Results and analysis of the mask lengths}
We next investigate how short speech clips extracted by different Boolean masks influence the performance of our proposed method via experiments conducted on the MLS14 data in EVAL17. We first evaluate the dual-mode LID models with random-location Boolean masks of different lengths. From Table~\ref{tab:mask_len_type}, it is not surprising that the model with the 3~s mask achieves the highest performance on 3~s test speech utterances, and the model with a longer mask achieves, in general, higher performance on 10~s and 30~s testing speech. However, we observed that the 6~s random mask shows lower performance on 30~s speech than the 5~s one. This may indicate that as the mimicked clips get longer, the short mode tends to pay more attention to the contextual dependencies than the local information, the enhancement resulted by better local information thus decreases. When the speech clips are as long as the full-length speech, the dual-mode LID system performs the worst, similar to the XSA-LID model. Consequently, the comparison between masks of different lengths implies that our proposed approach exhibits robustness to various speech durations without restricting to a specific Boolean mask length, although it is useful to note that an appropriate mask length can optimally improve the overall performance.
\subsubsection{Results and analysis of the mask location}
Apart from the window length, the position where the Boolean mask is located may also influence the performance of our model. In Table~\ref{tab:mask_len_type}, the label ``Fixed" denotes the Boolean mask extracting features of the speech in the first several seconds of the full-length speech signal while the label ``Random" denotes features of the speech clips being extracted by the mask from a random position of each utterance. 

With reference to Table~\ref{tab:mask_len_type}, compared to the 3~s fixed-location Boolean mask, the 3~s random-location mask exhibits higher performance on 3~s and 10~s test speech, while the former achieves modestly higher performance on 30~s test speech. These results are expected and conform with our proposition highlighted in Section~\ref{sec:booleanmask}. The speech clip extracted by the random-location Boolean mask can vary for each utterance in every epoch during training. Hence, it provides more clip-wise linguistic variability than the first 3~s speech clip. This clip-wise linguistic variability serves as data augmentation and leads to higher performance on short speech. 

On the other hand, these results suggest that the clip-wise lexical integrity plays an important role in reducing the difference between full-length speech and its short clip. Since the KD loss quantifies the difference between two distributions, a lower difference implies a lower KD loss contribution during training. Therefore, the dual-model LID model with the fixed-location Boolean mask suffers from lower performance degradation than that with the random-location mask, whereas it achieves lower performance on short utterances.

\section{Discussion}
\label{sec:discussion}
Considering that the data may not be balanced across different duration levels, the LID model should cater to the target data. Our work shows that the LID performance on long and short utterances can be affected by several factors. Therefore, our work can be adjusted to accommodate the target data once the statistic such as average duration is known. 

To promote the performance improvement on long speech, as described in Section~\ref{sec:kd}, a higher \(\alpha\) and \(\beta\) can be adopted in Equation~\ref{eq:loss_all} during training. In addition, an appropriate length of the Boolean mask is also proven to be effective in boosting the LID performance on long utterances. It is useful to note that although the fixed-location Boolean mask can also help the model achieve higher performance in long-utterance LID, the random-location mask is recommended due to its ability to achieve higher overall performance. 

To improve the model performance in the short-utterance LID task, a higher weight can be assigned to the KD loss in Equation~\ref{eq:loss_all}. A short Boolean mask is also helpful to achieve higher performance on short speech.
\section{Conclusion}
\label{sec:conclusion}
In this paper, we adapted the XSA model, which was originally proposed for language diarization, to the task of language identification, and proposed to employ a dual-mode framework with knowledge distillation to enhance the LID performance on various-duration speech. The experiments showed that our proposed dual-mode LID model achieves the highest overall performance on the MLS14 set of NIST LRE 2017 data with the random-location Boolean mask. We evaluated the contributions of the mimicked speech clips, the dual-mode framework, and the KD loss to the performance improvement, and discussed the influence of different Boolean masks in terms of the mask length and location. We found that the dual-mode framework and fixed-location provide more performance improvement on long utterances, while the performance on short utterances is attributed more to the KD loss and random-location mask.

\section{Acknowledgements}
This work was supported of the National Research Foundation, Singapore, under the Science of Learning programme (NRF2016-SOL002-011), and the Centre for Research and Development in Learning (CRADLE) at Nanyang Technological University, Singapore (JHU IO 90071537).

\bibliographystyle{IEEEtran}
\bibliography{odyssey_context}

%

\end{document}